\documentclass[prl,showpacs,preprintnumbers,twocolumn]{revtex4}
\usepackage{amssymb}
\usepackage{color}
\usepackage{graphicx}

\begin{document}

\title[Spin Wave]{Supermetallic conductivity in bromine-intercalated graphite}
\author{S. Tongay$^1$, J. Hwang$^{1,2}$, D. B. Tanner$^1$, H. K. Pal$^1$, D. Maslov$^1$, A. F. Hebard$^1$}
\email[Corresponding author:~]{afh@phys.ufl.edu}
\affiliation{$^1$Department of Physics, University of Florida, Gainesville FL 32611}
\affiliation{$^2$Department of Physics, Pusan National University, Busan 609-735, Republic of Korea}
\author{}
\keywords{}
\pacs{81.05.UW, 71.20.-b, 72.15.-v}

\begin{abstract}
Exposure of highly oriented pyrolytic graphite to bromine vapor gives rise to in-plane charge conductivities which increase monotonically with intercalation time toward values (for $\sim$6 at\% Br) that are significantly higher than Cu at temperatures down to 5~K. Magnetotransport, optical reflectivity and magnetic susceptibility measurements confirm that the Br dopes the graphene sheets with holes while simultaneously increasing the interplanar separation. The increase of mobility ($\sim 5 \times 10^4$ cm$^2$/V$\cdot$s at $T$=300~K) and resistance anisotropy together with the reduced diamagnetic susceptibility of the intercalated samples suggests that the observed supermetallic conductivity derives from a parallel combination of weakly-coupled hole-doped graphene sheets.\end{abstract}

\maketitle
\date{\today}

At the simplest level graphite can be thought of as an ordered stacking of weakly-coupled single graphene sheets. Delamination or deconstruction of graphite into isolated graphene sheets for experimental characterization by mechanical means\cite{novoselov666} has nucleated intense experimental and theoretical investigation into the electronic properties of this two-dimensional carbon allotrope \cite{castroneto109}. The presence of Dirac-like electronic excitations, an anomalous integer quantum Hall effect, and signature sensitivity to different types of disorder are but a few of the fascinating phenomena emerging from these studies. Bernal stacked graphite with an interplanar spacing $c = 3.4~\AA$ manifests properties that are precursors to the unusual behaviors associated with graphene and few-layer graphene. For example the presence of Dirac fermions near the H point in the Brillouin zone has been detected by angle-resolved photoemission spectroscopy\cite{zhou595}. 

Graphite intercalation compounds (GICs) have long been recognized as having unusual and sometimes surprising properties \cite{dresselhaus189}. In this study we take the approach of using bromine (Br) intercalents to simultaneously dope and separate the planes of graphite and thereby begin an approach to the limit where the interplanar coupling is sufficiently weak to assure that the resulting in-plane conductivity can be considered as the parallel contribution of relatively independent doped graphene sheets.  Since graphite is well compensated with approximately equal numbers of electron and holes giving a density on the order of $10^{-4}$ carriers per carbon atom\cite{brandt,du166601}, a small charge transfer between the intercalate and the adjacent carbon planes can result in a significant increase in free carriers per carbon. We find that random site interplanar doping with bromine to concentrations ($\leq 6$ at\%) gives rise to a pronounced decrease of the in-plane resistivity $\rho_{ab}$ to ``supermetallic''\cite{brandt243} values that are significantly lower than Cu over the temperature range 300~K$> T >$1.7~K. Hall and X-ray photoelectron 
spectroscopy (XPS) measurements confirm that the Br dopant acts like an acceptor, thus hole doping the graphene planes. Optical reflectance measurements confirm the supermetallic in-plane conductivity and further reveal a doping-induced increase of mobility and carrier density. The diamagnetic susceptibility decreases toward zero as would be expected for isolated graphene sheets\cite{mcClure}, and there is no evidence of diamagnetic screening that might be associated with superconducting fluctuations. At $T=5$~K the inferred sheet resistance per graphene plane of less than 1~$\Omega$ is significantly lower than reported for isolated graphene sheets either biased by an adjoining gate\cite{novoselov666} or doped with impurity atoms\cite{schedin652}.

The HOPG samples are cut from a single piece having $0.5^{\circ}$  mosaic spread and typically have dimensions on the order of 1-5~mm. The samples are exposed to Br gas at room temperature in closed tubes for various intercalation times and then removed and measured in a four-contact arrangement using a LR700 17~Hz resistance bridge at temperatures and fields along the $c$-axis in the ranges 5~K $< T <$ 300~K and -7~T $< B <$ 7~T respectively. In highly anisotropic samples, precautions are needed to assure uniform current density between voltage leads \cite{foley371}. We therefore used platelet (brick) shaped samples for the $\rho_{ab}$ ($\rho_{c}$) measurements while taking care to assure uniformly contacted current leads for each case. Results for the resistivities of a given sample are reproducible to 1\% and for twelve different samples cut from the same piece reproducible to  3\%. After each measurement additional intercalations could be made on the same sample. Weight uptake and volume increase, measured to an accuracy of $\sim$ 3\% give a good measure of the correlation between Br concentration and doping time $t_{Br}$.

\begin{figure}[t]
\includegraphics[angle=0,width=0.45\textwidth]{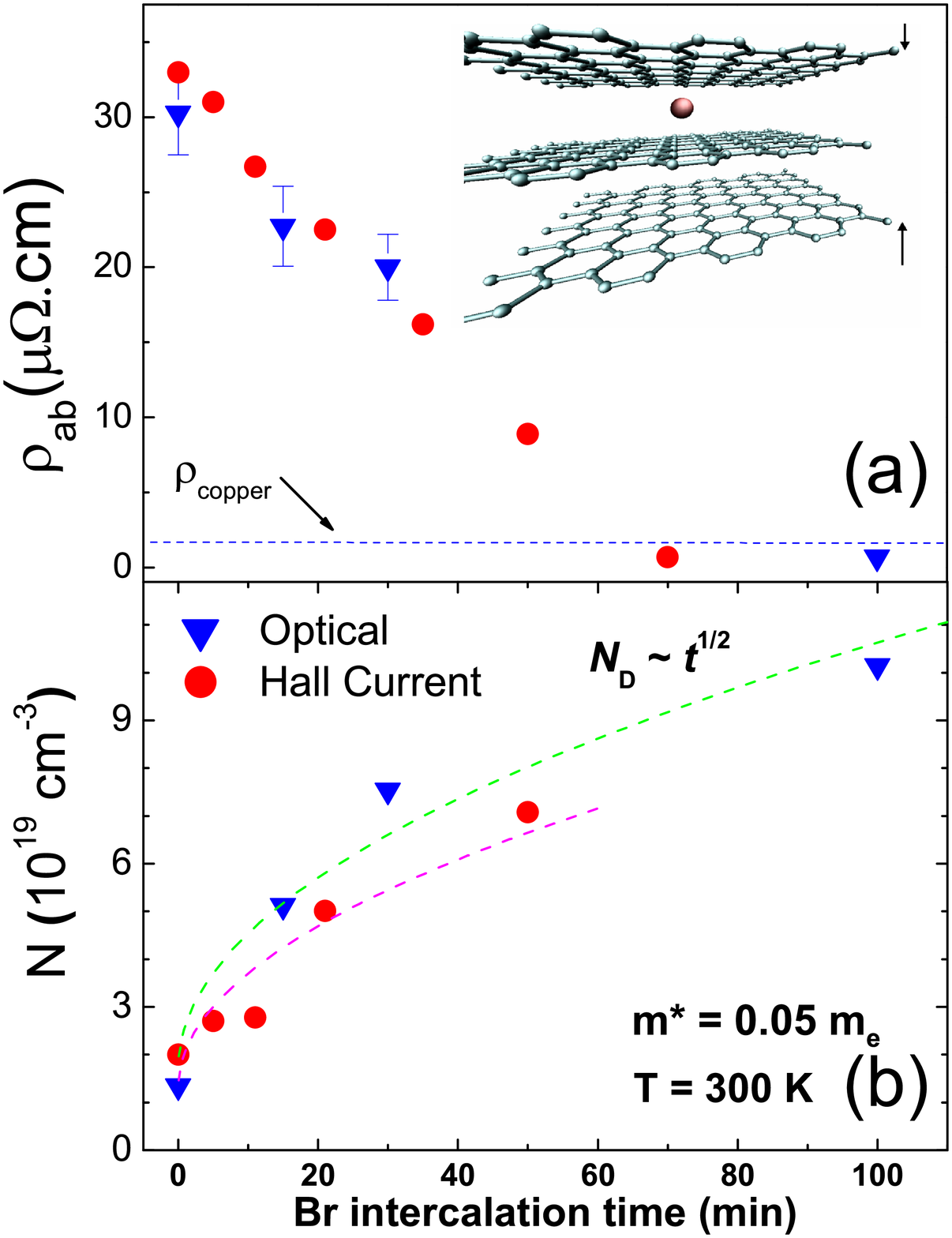}
\caption{\label{intercalation} Plots of the room temperature in-plane resistivity $\rho_{ab}$ (a) and carrier density $N$ (b) as a function of Br intercalation time. The solid red circles (blue triangles) in both panels are inferred from transport/Hall (optical reflectance) measurements. After 100 minutes $\rho_{ab}$ is reduced by a factor of 100 below its starting value to a resistivity that is a factor of five below that of copper (horizontal dashed line).}
\label{fig1}
\end{figure}

To further characterize the differences between pristine HOPG and the Br-doped samples, we also used X-ray diffraction (XRD), optical reflectance, Auger electron spectroscopy (AES), scanning electron microscopy (SEM) and x-ray photoelectron spectroscopy (XPS). SEM characterization did not detect any intercalation induced changes in surface morphology. The $\theta -2\theta$ XRD scans revealed only (00l) reflections which were shifted to lower angle (increased lattice spacing) with increased intercalation time and hence the concentration of Br between the graphene layers. This expansion is in agreement with an observed visual swelling of the sample. There were no additional reflections indicating staging\cite{dresselhaus189} or, equivalently, the ordering of the Br intercalants. Optical reflectance measurements were made at 300 K using a Bruker 113v Fourier spectrometer over the range 40-5000 cm$^{-1}$ (5-600 meV) and a Zeiss MPM 800 microspectrophotometer over 4000-40,000 cm$^{-1}$ (0.5-5 eV). The low frequency limit is set by the signal-to-noise and diffraction limitations of our small samples. Br LMM Auger peaks are observed to be located at 1396~eV, 1442~eV and 1476~eV for Br exposure times $\geq$ 30 minutes. The samples were repeatedly cleaved and remeasured to probe the Br concentration at different depths of the sample; the peak heights for a given sample remained constant to within 5~\%. The Br concentrations extracted from the Auger measurements agreed well with the weight-uptake/volume-expansion measurements.

XPS spectra of bromine doped HOPG samples were measured with a monochromatized Mg X-ray source with energies up to 1100~eV. Elemental percentage analyses were found to be consistent with AES and weight uptake measurements. The C$\:$1s electron binding energy measured relative to the Fermi level is observed to be at 284.5~eV for pristine and at 284.0~eV for the $t_{Br} =$70~min sample. At first sight, the reduction in the C$\:$1s binding energy by 0.5~eV contradicts the acceptor nature of Br, i.e the more positively charged carbon should, for fixed $E_{F}$, have a higher binding energy. However, similar trends/findings in donor (acceptor) type intercalants and associated increase (decrease) in C$\:$1s binding energy have been reported in the literature for different compounds\cite{Wertheim,Yan} and attributed to the change in $E_{F}$ before and after intercalation. In brominated HOPG, $E_{F}$ is negative and significantly larger in magnitude compared to pristine HOPG and the C$\:$1s binding energy is thus measured with respect to the lower $E_{F}$ of the hole-doped system. Accordingly, the increase in the C$\:$1s binding energy is more than compensated for by the decrease in $E_{F}$, giving an overall decrease in the C$\:$1s peak position as observed. The XPS method thus gives another way of estimating the change in Fermi level and implies a $\sim$-0.5~eV change in $E_{F}$ after hole doping to 6 at\% Br.
\begin{figure}[h]
\includegraphics[angle=0,width=0.5\textwidth]{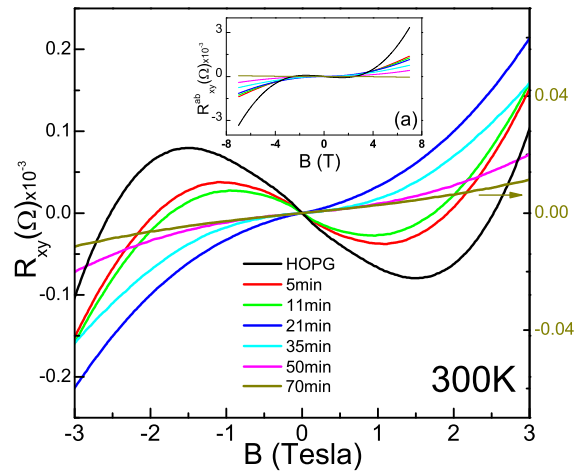}
\caption{\label{hall} Transverse resistance $\rho_{xy}$ as a function of perpendicular magnetic field $B$ for the indicated intercalation times. The inset shows the same data over a larger field range.}
\label{fig2}
\end{figure}

Figure \ref{intercalation}a shows the dependence of $\rho_{ab}$ on Br intercalation time at $T=300$~K. Although the initial linear dependence is not understood, we note that after 70~minutes, $\rho_{ab}$ appears to saturate at a value that is approximately a factor of five lower than the room temperature value ($1.7 ~\mu \Omega$cm) of copper indicated by the horizontal dashed line. Our interpretation of how the carrier density $N$, the scattering time $\tau$ and the effective mass $m^{\star}$ are affected by Br intercalation is based on using the Drude model in which the conductivity $\sigma$ of each contributing band is $\sigma = Ne^2 \tau /m^{\star} = N e \mu$, where the mobility $\mu = e \tau / m^{\star}$.

In Fig. \ref{hall} we show the evolution of the field ($B$) dependent transverse resistance $\rho_{xy}$ for intercalation times $t_{Br}$ ranging from $t_{Br} = 0$ (pristine HOPG) to $t_{Br} = 70$ min. The data for pristine HOPG are well fit by the expression for the two-band model\cite{du166601},
\begin{equation}  \label{rhoxy}
\frac{\rho_{xy}(B)}{e \rho_{xx}(0)^2} = \frac{\left( -n_e \mu_e^2 + n_h \mu_h^2 \right) B+
 \mu_e^2 \mu_h^2 \left( n_h - n_e \right) B^3 }{1 + e^2 \rho_{xx}(0)^2
 \mu_e^2 \mu_h^2 \left( n_h - n_e \right)^2 B^2}
\end{equation}
where $\rho_{xx} (0)$ is the $B = 0$ in-plane resistivity and the subscripts $e$ and $h$ refer respectively to the electron and hole bands. The obtained fitting parameters are $n_h = 2.0(1) \times 10^{19}$ cm$^{-3}$,
$n_e = 1.6(1) \times 10^{19}$ cm$^{-3}$,  $\mu_h$ =  4700(100) cm$^2$/V$\cdot$s and $\mu_e$ =  6800(100) cm$^2$/V$\cdot$s showing that our pristine HOPG is slightly hole doped with similar mobilities in each band. 
In graphite, Eq.~(\ref{rhoxy}) is applicable for $\mu B\gtrsim 1$,
when $\rho_{xx} (B)$ is quadratic in $B$\cite{pal}. At $T=300$~K, $\mu B=1$ corresponds to $B=0.3$~T in pristine graphite, so Eq.~(\ref{rhoxy}) describes most of the field range presented in Fig.~\ref{hall}. With increasing $t_{Br}$, the quadratic dependence of $\rho_{xx} (B)$ occurs over a decreasing field range, thus restricting the range of validity of Eq.~(\ref{rhoxy}). Constrained by this requirement, we extract a square-root dependence of $N$ on $t_{Br}$ shown in Fig \ref{intercalation}b. As seen in Fig.~\ref{hall}, the low-field slope, which is positive for $t_{Br} \geq$20~min. decreases with increasing Br concentration and becomes linear for the highest $t_{Br}$, thus indicating that Br is hole doping the graphene sheets with a carrier density $N = n_h$ dominated by holes.

\begin{figure}[h]
\includegraphics[angle=0,width=0.45\textwidth]{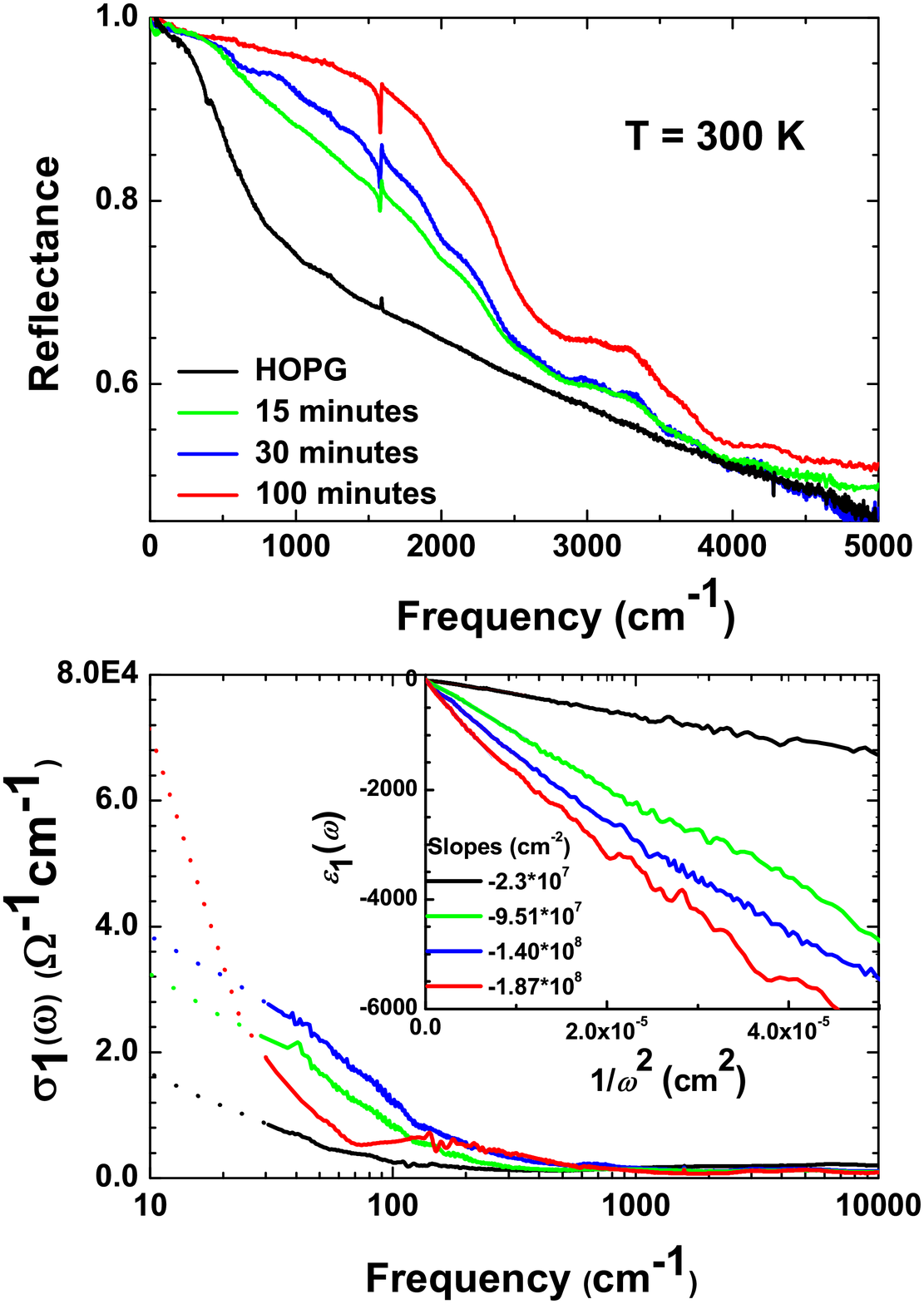}
\caption{\label{reflectance}
Infrared reflectance (a) and optical conductivity (b), $\sigma_1(\omega)$, at indicated intercalation times $t_{Br}$. Data below 35 cm$^{-1}$ in panel (b) are from the Drude-Lorentz fit to the reflectance.  Inset: Real part of the dielectric function versus 1/$\omega^2$. The slope, proportional to the plasma frequency squared, is a measure of the total carrier density.}
\label{fig3}
\end{figure}

With increasing $t_{Br}$ the far-infrared and midinfrared reflectance increase dramatically as shown in Fig.~\ref{reflectance}. The weak optical phonon at 1589 cm$^{-1}$ develops a Fano lineshape in reflectance, implying that there is metallic behavior out to these energies in the strongly intercalated samples.
This expectation is borne out by a Kramers-Kronig analysis\cite{wooten}. We used a Drude extrapolation at low frequencies and a power law behavior at high frequencies, with the results shown in Fig.~\ref{reflectance}b. Note that the optical conductivity curves include frequencies (dashed lines) where we used the extrapolation; the good agreement with the dc conductivity in Fig.~\ref{intercalation}a and the good fits to the reflectance provide confidence that the behavior is as shown. As the doping proceeds, the low-frequency conductivity increases and the spectral weight (the area under the curve) increases significantly (Fig.~\ref{reflectance}b). The full width of this Drude-like peak represents the carrier scattering rate $\tau^{-1}$which at the highest doping is decreased by a factor of five relative to the pristine sample.

From the sum rule on the optical conductivity, we can relate the low-energy spectral weight to the carrier density. The spectral weight also affects the real part of the dielectric function, $\epsilon_1 (\omega)$, which for free carriers follows $\epsilon_1(\omega) = 1 - \omega^{2}_{p}/\omega^2$, where $\omega^{2}_{p} = 4\pi Ne^2/m^\star$ is the plasma frequency. The inset shows $\epsilon_1 (\omega)$ plotted vs 1/$\omega^2$; the slopes of these plots give the carrier density, which is seen to increase by more than a factor of eight with increasing $t_{Br}$. That the curves are straight lines implies strongly that a free-carrier (metallic) picture of the low-energy electrodynamics is an accurate view of the intercalated graphite. The dc resistivities inferred from the Drude fit to the infrared reflectance measurements are shown as blue triangles in Fig.~\ref{intercalation}a. Because the reflectance measurements are made without placing electrical contacts on the sample, the inferred dc conductivities of Fig.~\ref{reflectance}b give independent confirmation of the supermetallic conductivity inferred from transport measurements.

\begin{figure}[t]
\includegraphics[angle=0,width=0.55\textwidth]{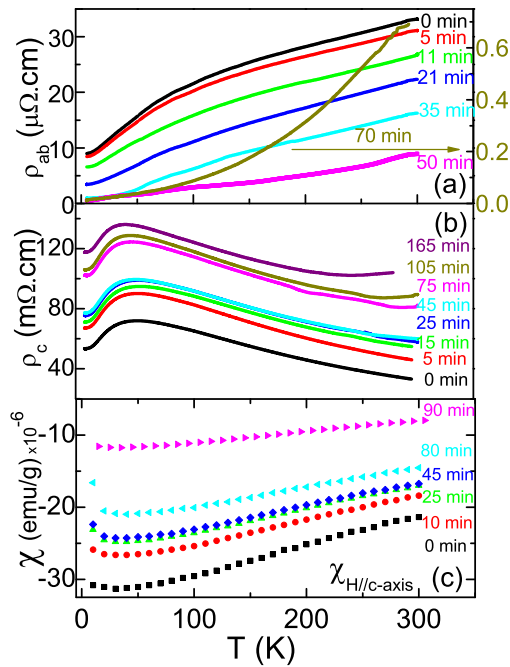}
\caption{\label{Tdependence} Temperature dependence of (a) $\rho_{ab}$, (b) $\rho_c$ and (c) magnetic susceptibility $\chi_c$ at the indicated intercalation times. The right hand axis of panel (a) is an expanded scale for the $t_{Br}$=70~min. curve}
\label{fig4}
\end{figure}

The temperature dependent resistivity data of panels (a) and (b) of Fig.~\ref{Tdependence} show that the resistivity scales for $\rho_{ab}$ ($\rho_c$) decrease (increase) as $T$ is reduced from 300~K to 5~K. The positive curvature for samples with $t_{Br}\geq$70~min. is consistent with the notion that the doping is sufficient to guarantee $E_F \gg k_BT$ in contrast to pristine HOPG and the lightly doped samples where $E_F \sim$300~K and significant variation of $N$ with $T$ leads to a negative curvature of $\rho_{ab} (T)$. The ratio $\rho_{ab}$(300K)/$\rho_{ab}$(5K) = 47 for the sample with $t_{Br}$=70~min. is higher by more than a factor of 10 than the same ratio (4.0) for pristine HOPG. In contrast to $\rho_{ab}$, $\rho_c$ increases with increasing $t_{Br}$ due to the presence of Br intercalents acting like a negative pressure pushing the planes apart (see schematic inset of Fig.~\ref{intercalation}a), thereby resulting in decreased interplanar tunneling. With the application of positive pressure the interplanar spacing decreases and there is a corresponding decrease in $\rho_c$\cite{uher}. The hole doping of the planes (decreasing $\rho_{ab}$) concomitant with an increasing interplanar spacing (increasing $\rho_c$) leads to an anisotropy ratio $\rho_c / \rho_{ab}$ at 5~K approaching 10$^7$, a factor of 1000 greater than measured for pristine HOPG at the same temperature. 

Panel (c) of Fig.~\ref{Tdependence} shows that the temperature dependent dc diamagnetic susceptibility $\chi$ (field parallel to \textit{c}-axis) decreases with increasing $t_{Br}$. Our room temperature value $\chi$ =~-21.3$\times 10^{-6}$~emu/g for pristine graphite, which is in good agreement with previous experiments\cite{brandt,mcClure}, decreases by a factor of three for $t_{Br}$=90~min. This decrease in $\chi$ with increased hole doping of the graphene planes is qualitatively understood by realizing that as the Fermi energy moves away from the neutrality point ($n_h = n_e$) of compensated pristine graphite, the cyclotron mass $m_{c}^{\star}$ increases and $\chi \propto 1/ m_{c}^{\star}$ decreases, approaching the limit of exponentially weak diamagnetism for single-plane Dirac fermions. Importantly, there is no signature of superconductivity, which, if associated with the giant conductivity, would become manifest as an \textit{increase} in diamagnetism at some characteristic temperature.

Graphite intercalation compounds in the dilute limit are well known to exhibit enhancements of room-temperature conductivity  which, with increasing intercalant concentration, saturate to modest values $\sim$10 times the pristine value\cite{dresselhaus3180}. The surprizing and unexpected result presented here is that for uniformly dispersed non-staged Br dopants at relatively low concentrations near 5-6~at\%, the conductivity can justifiably be referred to as ``supermetallic''. To make comparisons to single-layer graphene, we convert our $\rho_{ab}$ measurements to resistance per square $R_g$ of each graphite layer and see that $R_g$ near 1000~$\Omega$ for HOPG at 300~K decreases to less than 0.5~$\Omega$ at 5~K for intercalated samples with $t_{Br}$=70~min. To our knowledge there are no reports of such a low $R_g$ for graphene. 

The experimental results raise many questions, e.g., since doping introduces disorder, why does the conductivity increase? At high doping level, most of the carriers (holes) come from negatively charged acceptors, so we have a system in which the number of electrons is equal to the number of scattering centers. A priori, this should be a dirty system. However, since the Fermi energy increases with doping, the 2D Rutherford scattering cross-section decreases: $A_R\sim e^2/E_F$. Assuming that the 2D scaling $E_F\propto n_a$, where $n_a=n_hc$  is the number density of acceptors per layer, and estimating the mean free path as $\ell =1/n_a A_R$, we arrive at a simple result for the parameter $k_F\ell$, which characterizes purity of a material: $k_F\ell\sim 1/r_s$, where $r_s$ is the average inter-carrier distance measured in Bohr radii. Already in pristine graphite, $r_s\approx 1$ \cite{gutman} and it decreases further with doping. Therefore, the material stays \lq\lq clean\rq\rq\ despite a one-to-one ratio of scatterers to carriers.

In addition, the mobility of Br-intercalated HOPG at room temperature, confirmed by optical measurements, is 50,000~cm$^2$/V$\cdot$sec, a factor of 5 higher than pristine HOPG. From transport data at 5~K we find $\mu_h \sim 10^6$ cm$^2$/V$\cdot$ s but consider this with some reservations in the absence of confirming optical data. Although we have not reached the limit where the interplanar coupling is sufficiently low to consider our intercalated graphite as an ordered stack of isolated graphene sheets each of which is dominated by Dirac fermions, we believe our results illustrate the emergence of intriguing phenomenology at the graphite/graphene boundary accessed by intercalation.

We thank D. Arenas, H-P Cheng, V. Craciun, J. E. Fischer and E. Lambers for useful discussions. This research was supported by the NSF and DOE under grant numbers DMR-0704240(AFH) and DE-FG02-02ER45984(DBT).

\end{document}